\PassOptionsToPackage{table}{xcolor}
\documentclass[sigplan,10pt]{acmart}

\renewcommand\footnotetextcopyrightpermission[1]{}

\usepackage{tikz}
\usetikzlibrary{positioning}
\usepackage{url}
\usepackage{array}
\usepackage{braket}
\usepackage{subfig}
\usepackage{amsmath}
\usepackage{balance}
\usepackage{environ}
\usepackage{physics}
\usepackage{siunitx}
\usepackage{tabularx}
\usepackage{listings}
\usepackage{arydshln}
\usepackage{makecell}
\usepackage{quantikz}
\usepackage{colortbl}
\usepackage{enumitem}
\usepackage{amsfonts}
\usepackage{booktabs}
\usepackage{graphicx}
\usepackage{textcomp}
\usepackage{multirow}
\usepackage{algorithm}
\usepackage{tcolorbox}
\usepackage{algorithmic}
\usepackage[normalem]{ulem}

\definecolor{codecomment}{RGB}{0,150,0}
\definecolor{codekeyword}{RGB}{0,92,197}
\definecolor{codeassert}{RGB}{210,0,0}
\definecolor{codestring}{RGB}{180,0,160}

\lstdefinelanguage{QuantumPseudo}{
  morekeywords={h,cx,rz},
  sensitive=true,
  morecomment=[l]{\#},
  morestring=[b]"
}

\lstdefinestyle{quantumcode}{
  language=QuantumPseudo,
  basicstyle=\ttfamily\footnotesize,
  keywordstyle=\color{codekeyword}\bfseries,
  commentstyle=\color{codecomment}\itshape,
  stringstyle=\color{codestring},
  numbers=left,
  numberstyle=\tiny\color{codecomment},
  stepnumber=1,
  numbersep=6pt,
  showstringspaces=false,
  frame=none,
  columns=fullflexible,
  keepspaces=true,
  alsoletter={_},
  escapeinside={(*@}{@*)},
  emph={assert_classical, assert_superposition, assert_entanglement, Z,X},
  emphstyle=\color{codeassert}\bfseries
}

\newcommand{\blackcircle}[1]{%
  \tikz[baseline=(char.base)]{
    \node[shape=circle, draw=black, fill=black, text=white, inner sep=1pt, font=\small\bfseries] (char) {#1};
  }%
}

\newcommand{\sol}{SynDebug}

\newif\ifhighlight
\highlighttrue

\begin{document}

\title[\sol{}: A Framework to Debug Fault-Tolerant Quantum Computing Programs]{\sol{}: A Framework to Debug Logic Bugs in Fault-Tolerant Quantum Computing Programs}

\author{Nicholas S. DiBrita}
\affiliation{%
  \institution{Rice University}
  \city{Houston}
  \state{TX}
  \country{USA}}

\author{Jack Lange}
\affiliation{%
  \institution{Oak Ridge National Laboratory}
  \city{Oak Ridge}
  \state{TN}
  \country{USA}}

\author{Tirthak Patel}
\affiliation{%
  \institution{Rice University}
  \city{Houston}
  \state{TX}
  \country{USA}}

\renewcommand{\shortauthors}{DiBrita et al.}

\begin{abstract}

Fault-tolerant quantum computing (FTQC) introduces substantial systems-level complexity through quantum error correction (QEC), making debugging FTQC programs challenging. Existing quantum debugging approaches primarily target logical-level circuits and do not naturally support debugging within realistic FTQC execution flows. We present \sol{}, the first systems-level debugging framework for FTQC programs. \sol{} introduces lightweight assertion-based debugging primitives that operate directly within encoded FTQC infrastructures while avoiding additional logical surface area and minimizing temporal overhead. We evaluate \sol{} across multiple QEC code families, code distances, and encoded workloads, including state preparation, entanglement validation, and Shor-style programs. Our results demonstrate that \sol{} enables practical FTQC debugging with an average 1.8\% additional circuit depth overhead (0\% in most cases) and no measurable increase in the logical error rate, showing that systems-level debugging for large-scale FTQC applications is both feasible and practical.

\end{abstract}

\maketitle

\section{Introduction}
\label{sec:introduction}

\begin{figure}
    \centering
    \includegraphics[width=0.99\linewidth]{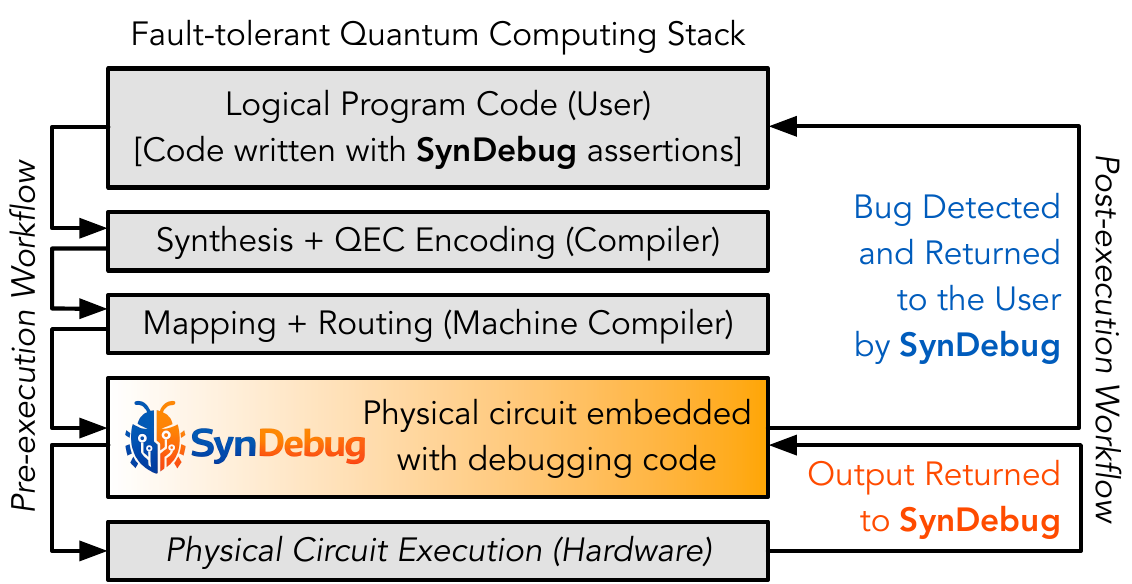}
    \vspace{-3mm}
    \caption{\sol{} is a systems-level framework for debugging fault-tolerant quantum computing programs.}
    \label{fig:positioning}
    \vspace{-1mm}
\end{figure}

Quantum computing is steadily advancing toward fault-tolerant quantum computing (FTQC), enabling the execution of deep, large-scale quantum workloads for applications in scientific simulation, optimization, cryptography, machine learning, and chemistry \cite{preskill2023quantum,gill2025quantum}. Achieving FTQC requires sophisticated quantum error correction (QEC) infrastructures that continuously detect and correct physical faults to maintain the integrity of logical quantum states \cite{maurya2026case,google2025quantum,katabarwa2024early}. These systems introduce substantial implementation complexity through functions such as logical encoding, physical mapping, syndrome extraction, lattice surgery, transversal operations, and ancilla preparation \cite{turner2026scalable,marton2025lattice}. Consequently, debugging FTQC programs becomes substantially more difficult than debugging conventional software or even noisy intermediate-scale quantum (NISQ) programs.

Many quantum software bugs arise from mistakes in program logic rather than failures in the underlying hardware. These include reversed gate ordering, incorrect gate operands, swapped control and target qubits, missing gates, duplicated gates, and incorrect parameterization of quantum operations \cite{huang2019statistical,veganzones2025assertsq,kole2026assertion}. While such issues are often straightforward to identify in classical systems through tracing, print statements, memory inspection, or step-by-step execution, these approaches do not naturally extend to quantum programs.

This is because, first, observing the state of a quantum variable collapses the underlying computation, preventing debugging strategies that rely on tracing the state \cite{patel2020ureqa}. Second, quantum measurements reveal only partial information about the underlying state \cite{swierkowska2026eccentric}. Third, many debugging approaches would require exact or approximate quantum simulation, which scales exponentially with program size and rapidly becomes infeasible for realistic FTQC systems \cite{patel2026selective,tanji2026quantum}. Fourth, quantum programs inherently produce probabilistic outputs and uncertain intermediate states, making it difficult to distinguish expected stochastic variation from actual program bugs \cite{huang2019statistical,li2025statistical}. Lastly, FTQC systems introduce additional complexity through the functions mentioned above, all of which substantially complicate debugging at the system level. These challenges make debugging FTQC quantum programs at once both critical and non-trivial.

\vspace{2mm}

\noindent\textbf{The Problem.} Existing debugging approaches for quantum computing primarily target logical circuits in the NISQ regime~\cite{li2022exploiting,li2020projection,liu2020quantum,liu2021systematic,kole2026assertion,veganzones2025assertsq,oldfield2025bloch,li2025statistical,huang2019statistical,rovara2025framework2,rovara2025automatically}. These methods typically assume direct access to relatively small logical circuits for NISQ settings. In FTQC hardware, however, logical circuits must ultimately be translated into complex FTQC implementations for execution. Consequently, existing approaches do not naturally support debugging in this regime.

\vspace{2mm}

\noindent\textbf{The Gap.} Existing approaches provide important foundations for quantum software reliability, but extending these techniques to FTQC systems introduces several constraints. (1) FTQC systems involve diverse, rapidly evolving hardware technologies and QEC methodologies \cite{zhu2025quantum,fang2026bridging}, requiring debugging frameworks that remain generalizable. (2) FTQC systems already incur extremely large physical qubit overheads due to QEC encoding \cite{trochatos2026traceq,maurya2026case,kim2026architecting}, making it undesirable for debugging infrastructures to require additional logical surfaces or dedicated debugging qubits. (3) Additional debugging gates and measurements may increase circuit depth and runtime. (4) Debugging techniques must remain scalable to large encoded systems without relying on infeasible approaches such as full-state tomography or exponentially scaling simulation, while still remaining useful.

\vspace{2mm}

\noindent\textbf{Design Principles.} \sol{}\footnote{\sol{}'s name reflects how debugging is performed through QEC syndromes and encoded execution structures in FTQC implementations. This work is published in the proceedings of the 2026 ACM SIGOPS Annual Technical Conference (SIGOPS ATC 2026).} follows five principles (Section~\ref{sec:reqs}): (1) generality across hardware and QEC methods through stabilizer-code abstractions, (2) no debugging-specific logical surface, (3) minimal temporal perturbation, (4) scalability through lightweight logical observables, and (5) a developer-facing logical API. More broadly, developers specify logical invariants while \sol{} handles their physical realization, and probabilistic behavior is assessed by comparing observed and expected outcome distributions using sound statistical tests. We also rely on a standard fact of quantum measurement, which is not our contribution: measuring an observable of which the state is an eigenstate leaves the state unchanged. Consequently, an assertion on a register expected to be in a basis state can allow execution to continue, while an assertion on a superposed or entangled register destroys the state and acts as a checkpoint.

\vspace{2mm}

% \footnote{\sol{}'s name reflects how debugging is performed through QEC syndromes and encoded execution structures in FTQC implementations.}

\noindent\textbf{Our Solution.} \sol{} is a systems-level debugging framework for FTQC programs that enables lightweight assertion-based validation directly within encoded execution flows, as shown in Figure~\ref{fig:positioning}. \sol{} introduces three user-facing debugging primitives, (1) \texttt{assert\_classical}, (2) \texttt{assert\_} \texttt{superposition}, and (3) \texttt{assert\_entanglement}, which allow developers to validate logical states, superposition structure, and entanglement relationships during FTQC execution. To remain practical for large-scale fault-tolerant systems, \sol{} is designed to operate across diverse stabilizer-code families while avoiding additional logical qubits and minimizing debugging overhead.

\vspace{2mm}

\noindent\textbf{The contributions of this work are as follows:}

\begin{itemize}[leftmargin=*]

    \item We introduce the first systems-level debugging framework for FTQC programs with user-facing APIs that integrates assertion-based debugging directly into encoded execution flows while remaining compatible with evolving QEC methods and hardware technologies.

    \item We design the \sol{} framework that avoids requiring additional logical surface area or dedicated debugging qubits, enabling compatibility with constrained FTQC layouts and encoded patch-based execution.

    \item We implement practical debugging primitives including \texttt{assert\_classical}, \texttt{assert\_superposition}, and \texttt{assert \_entanglement} with monitoring and instrumentation methods while integrating with existing FTQC libraries including Stim~\cite{stim}, the Munich Quantum Toolkit (MQT) QECC module~\cite{mqt,mqtstateprep}, and Qiskit \cite{qiskit}.

    \item We evaluate \sol{} across multiple QEC codes, including codes such as Color, Surface, Tetrahedral, Golay, and Steane, for different code distances $d \in \{3,5,7\}$.

    \item We demonstrate \sol{} through multiple FTQC case studies, including logical state preparation, encoded entanglement validation, and a Shor-style arithmetic workload demonstration, showing that practical FTQC debugging can be achieved with low additional overhead: an average 1.8\% increase in circuit depth (0\% increase in most cases), thus negligibly affecting the logical error rate.

    \item We open-source all of \sol{}'s source code, tools, datasets, and experimental artifacts to support reproducibility and future FTQC debugging research: \\\textit{\url{https://github.com/positivetechnologylab/SynDebug}}.

\end{itemize}
\section{Brief and Relevant Background}
\label{sec:background}

Quantum computing has the potential to provide substantial computational advantages for applications in scientific simulation, chemistry, optimization, machine learning, cryptography, and large-scale search \cite{preskill2023quantum,gill2025quantum}. However, quantum hardware will remain highly error-prone due to decoherence, imperfect control operations, measurement noise, crosstalk, and environmental interactions \cite{giortamis2025qos,tao2025quantum,patel2020ureqa}. As a result, executing deep quantum programs reliably requires FTQC, in which quantum computations are protected by QEC. Since \sol{} operates in this FTQC setting, we provide a brief, relevant background to introduce the challenges of debugging encoded quantum programs. %\rev{As a primer, qubits can be in a \textit{basis state} ($\ket{0}$ or $\ket{1}$), in a \textit{superposition} (e.g., $\ket{+}$ or $\ket{-}$), or \textit{entangled} (e.g., a Bell pair, whose qubits yield matching outcomes). A measurement in the $Z$ (or $X$) basis leaves the states $\ket{0}$ and $\ket{1}$ (or $\ket{+}$ and $\ket{-}$), called its \textit{eigenstates}, undisturbed, but it destroys any other superposition. Thus, programs are executed many times (\textit{shots}) to observe an outcome distribution.}

\subsection{Fault-tolerant Quantum Computing (FTQC)}

FTQC refers to the capability of executing large and deep quantum programs while maintaining computational fidelity through the continuous detection and correction of errors \cite{kim2026architecting}. Unlike current NISQ systems, which are limited by short coherence times and high error rates, FTQC systems aim to sustain computation indefinitely by actively correcting faults throughout execution. Achieving FTQC requires substantially more infrastructure than simply executing a logical quantum circuit. In practice, a logical program must first be synthesized, encoded into a QEC representation, mapped onto a physical hardware topology, routed according to hardware constraints, and executed in tandem with continuous syndrome extraction, decoding, and correction pipelines. Consequently, the physical FTQC implementation becomes significantly more complex than the original logical circuit.

FTQC systems are built on top of QEC codes, which encode logical qubits into ensembles of physical qubits. These encoded representations enable the system to potentially detect and correct errors without directly observing the quantum state. The resulting execution flow, therefore, involves repeated stabilizer measurements, ancilla preparation, syndrome decoding, and corrective operations throughout program execution \cite{turner2026scalable,marton2025lattice,sales2025experimental}. These additional layers of abstraction and runtime orchestration are what make debugging FTQC programs significantly more challenging than debugging either classical software or NISQ circuits.

\subsection{Quantum Error Correction (QEC) Codes}

QEC codes are the fundamental mechanism that enables FTQC systems to tolerate noise and execute long-running quantum computations reliably \cite{huang2025efficient,fang2024symbolic}. Similar to classical error correction, QEC protects information by encoding a smaller set of logical bits into a larger set of physical bits to introduce redundancy. In the quantum setting, however, the problem is substantially more difficult because quantum systems must simultaneously protect against both bit-flip and phase errors while preserving fragile quantum superpositions and entanglement \cite{zeng2026quantum,rodriguez2024witnessing}.

Most practical FTQC architectures are based on a family of codes known as \textit{stabilizer codes} \cite{poulin2005stabilizer,dauphinais2024stabilizer}. Stabilizer codes define a set of observables, called stabilizers, whose measurements reveal whether an error has occurred without collapsing the underlying logical quantum information. These stabilizer measurements produce \textit{syndromes}, which classical decoders interpret to infer and correct likely faults.

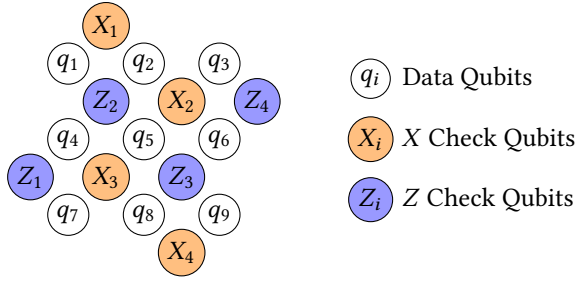
\begin{figure}[t]
\centering
\begin{tikzpicture}[scale=1]

% styles
\tikzstyle{data}=[circle,draw,fill=white,inner sep=1.6pt]
\tikzstyle{xcheck}=[circle,draw,fill=orange!50,inner sep=1.6pt]
\tikzstyle{zcheck}=[circle,draw,fill=blue!40,inner sep=1.6pt]

% DATA QUBITS
\node[data] (d1) at (0,2) {$q_1$};
\node[data] (d2) at (1,2) {$q_2$};
\node[data] (d3) at (2,2) {$q_3$};

\node[data] (d4) at (0,1) {$q_4$};
\node[data] (d5) at (1,1) {$q_5$};
\node[data] (d6) at (2,1) {$q_6$};

\node[data] (d7) at (0,0) {$q_7$};
\node[data] (d8) at (1,0) {$q_8$};
\node[data] (d9) at (2,0) {$q_9$};

% X-CHECKS (centered between rows)
\node[xcheck] at (0.5,2.5) {$X_1$};
\node[xcheck] at (1.5,1.5) {$X_2$};
\node[xcheck] at (0.5,0.5) {$X_3$};
\node[xcheck] at (1.5,-0.5) {$X_4$};

% Z-CHECKS (edges)
\node[zcheck] at (-0.5,0.5) {$Z_1$};
\node[zcheck] at (0.5,1.5) {$Z_2$};
\node[zcheck] at (1.5,0.5) {$Z_3$};
\node[zcheck] at (2.5,1.5) {$Z_4$};

% Legend
\node[data] at (4,1.8) {$q_i$};
\node[right] at (4.3,1.8) {Data Qubits};

\node[xcheck] at (4,1.0) {$X_i$};
\node[right] at (4.3,1.0) {$X$ Check Qubits};

\node[zcheck] at (4,0.2) {$Z_i$};
\node[right] at (4.3,0.2) {$Z$ Check Qubits};

\end{tikzpicture}
\vspace{-2mm}
\caption{Layout of an example QEC patch. This figure shows the distance-3 [[9, 1, 3]] rotated surface code.}
\label{fig:d3_patch}
\vspace{-1mm}
\end{figure}

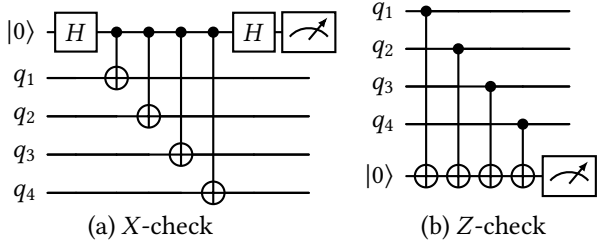
\begin{figure}[t]
\centering

\begin{minipage}{0.48\columnwidth}
\centering
\begin{quantikz}[row sep=0.18cm,column sep=0.1cm]
\lstick{$\ket{0}$} & \gate{H} & \ctrl{1} & \ctrl{2} & \ctrl{3} & \ctrl{4} & \gate{H} & \meter{} \\
\lstick{$q_1$} & \qw & \targ{} & \qw & \qw & \qw & \qw & \qw \\
\lstick{$q_2$} & \qw & \qw & \targ{} & \qw & \qw & \qw & \qw \\
\lstick{$q_3$} & \qw & \qw & \qw & \targ{} & \qw & \qw & \qw \\
\lstick{$q_4$} & \qw & \qw & \qw & \qw & \targ{} & \qw & \qw
\end{quantikz}

(a) $X$-check

\end{minipage}
\hfill
\begin{minipage}{0.48\columnwidth}
\centering
\begin{quantikz}[row sep=0.35cm,column sep=0.1cm]
\lstick{$q_1$} & \ctrl{4} & \qw & \qw & \qw & \qw  \\
\lstick{$q_2$} & \qw & \ctrl{3} & \qw & \qw & \qw  \\
\lstick{$q_3$} & \qw & \qw & \ctrl{2} & \qw & \qw  \\
\lstick{$q_4$} & \qw & \qw & \qw & \ctrl{1} & \qw \\
\lstick{$\ket{0}$} & \targ{} & \targ{} & \targ{} & \targ{} & \meter{}
\end{quantikz}
\vspace{1mm}

(b) $Z$-check

\end{minipage}

\vspace{-3mm}

\caption{$X$-check and $Z$-check stabilizer measurements from the surrounding data qubits in a rotated surface code.}
\label{fig:xzcheck}
\vspace{-1mm}
\end{figure}

QEC codes are commonly described using the notation $[[n,k,d]]$, where $n$ represents the number of physical qubits used by the code, $k$ represents the number of encoded logical qubits, and $d$ represents the code distance \cite{chen2021exponential}. Intuitively, the code distance determines how many physical errors the code can tolerate before logical information becomes corrupted. For example, the rotated surface code shown in Figure~\ref{fig:d3_patch} corresponds to a distance-3 $[[9,1,3]]$ code consisting of nine data qubits and additional stabilizer ancilla qubits.

Modern QEC systems repeatedly execute syndrome extraction circuits throughout computation. Figure~\ref{fig:xzcheck} illustrates $X$-check and $Z$-check circuits in a rotated surface code. In these circuits, ancilla (i.e., check) qubits interact with neighboring data qubits and are subsequently measured to extract syndrome information indicating whether an error likely occurred. These stabilizer measurements must be performed continuously during execution \cite{ravi2023better,maurya2026case}.

\subsection{Types and Varieties of QEC Codes}

The QEC landscape is both large and rapidly evolving, with different code families designed to optimize for distinct hardware constraints, connectivity assumptions, decoding complexity, and fault-tolerance properties \cite{ravi2023better,stein2025hetec}. Different quantum hardware technologies naturally favor different QEC methodologies due to variations in qubit connectivity, native operations, and coherence times \cite{fang2026bridging,viszlai2023architecture}.

Stabilizer codes are a type of QEC codes that are the quantum equivalent of classical linear block codes. Instead of traditional parity checks, they use a set of commuting Pauli operators to detect errors without destroying the quantum state \cite{pontolillo2025qucheck,dauphinais2024stabilizer}. One major class of Stabilizer QEC codes is \textit{CSS codes} (Calderbank-Shor-Steane codes) \cite{calderbank1996good,steane1996multiple}. CSS codes construct quantum error correction schemes by separately handling bit-flip ($X$) and phase-flip ($Z$) errors using two related classical linear codes. This separation simplifies stabilizer construction and decoding while enabling efficient implementations of several logical operations. Many important QEC families, including surface codes and color codes, can be expressed within the broader CSS framework.

Surface codes are among the most widely studied CSS approaches because they require only nearest-neighbor interactions and map naturally onto two-dimensional superconducting and neutral-atom architectures \cite{google2025quantum,fang2026bridging}. 

On the other hand, color codes provide alternative stabilizer structures with favorable transversal gate properties \cite{kubica2015universal}. This diversity creates an important systems challenge for debugging infrastructures. A practical FTQC debugging framework cannot assume a single hardware architecture, stabilizer layout, or QEC methodology because the underlying quantum computing ecosystem continues to evolve rapidly. Consequently, debugging mechanisms must remain sufficiently general to operate across multiple QEC code families and hardware technologies while still exposing meaningful debugging abstractions to developers.
\section{Motivation for \sol{}}
\label{sec:motivation}

\begin{figure}
    \centering
    \includegraphics[width=0.99\linewidth]{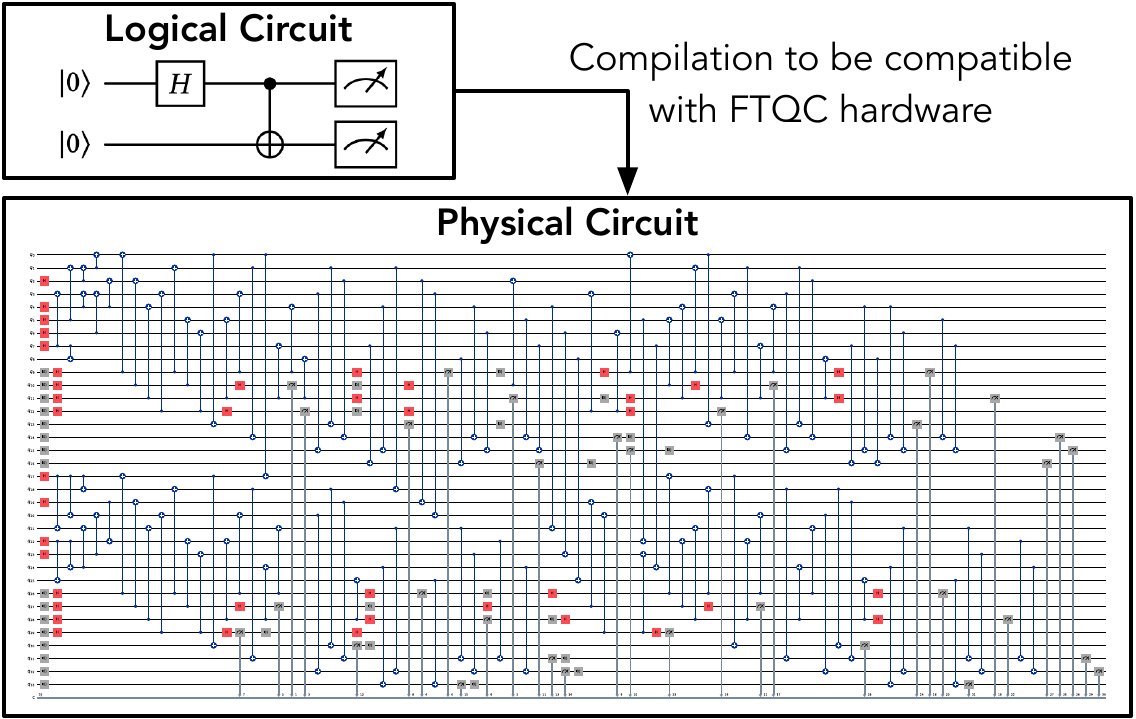}
    \vspace{-3mm}
    \caption{A simple logical two-qubit quantum circuit can become complex with the physical FTQC implementation. In this example, a Bell-pair circuit is implemented with a distance-3 surface code: [[9, 1, 3]]. While the specific gates and qubits in the physical circuit are illegible, the visual demonstrates its size and complexity.}
    \label{fig:bell_pair}
    \vspace{-1mm}
\end{figure}

\begin{algorithm}[t]
\caption{Correct vs. Buggy Implementation}
\label{code:example}
\vspace{-2mm}
\begin{tcolorbox}[boxrule=0.2mm, colframe=black, colback=blue!5!white]
\begin{lstlisting}[style=quantumcode]
# Correct implementation to prepare a Bell Pair
# Initialize register as |+>|0>
code = CSSCode.from_code_name("Surface", 7)
prog = QecProgram(code, 2, "p0")
# Add CNOT gadget from logical 0 to 1
transversal_cnot((*@prog@*), 0, 1)
# Quantum gate operations using Bell pair...
\end{lstlisting}
\end{tcolorbox}
\vspace{-4mm}
\begin{tcolorbox}[boxrule=0.2mm, colframe=black, colback=blue!5!white]
\begin{lstlisting}[style=quantumcode]
# Buggy implementation to prepare a Bell Pair
# Initialize register as |+>|0>
code = CSSCode.from_code_name("Surface", 7)
prog = QecProgram(code, 2, "p0")
# Add CNOT gadget with incorrect direction
transversal_cnot(prog, 1, 0)
# Quantum gate operations using Bell pair...
\end{lstlisting}
\end{tcolorbox}
\vspace{-2mm}
\end{algorithm}

\begin{figure*}
    \centering
    \includegraphics[width=0.99\linewidth]{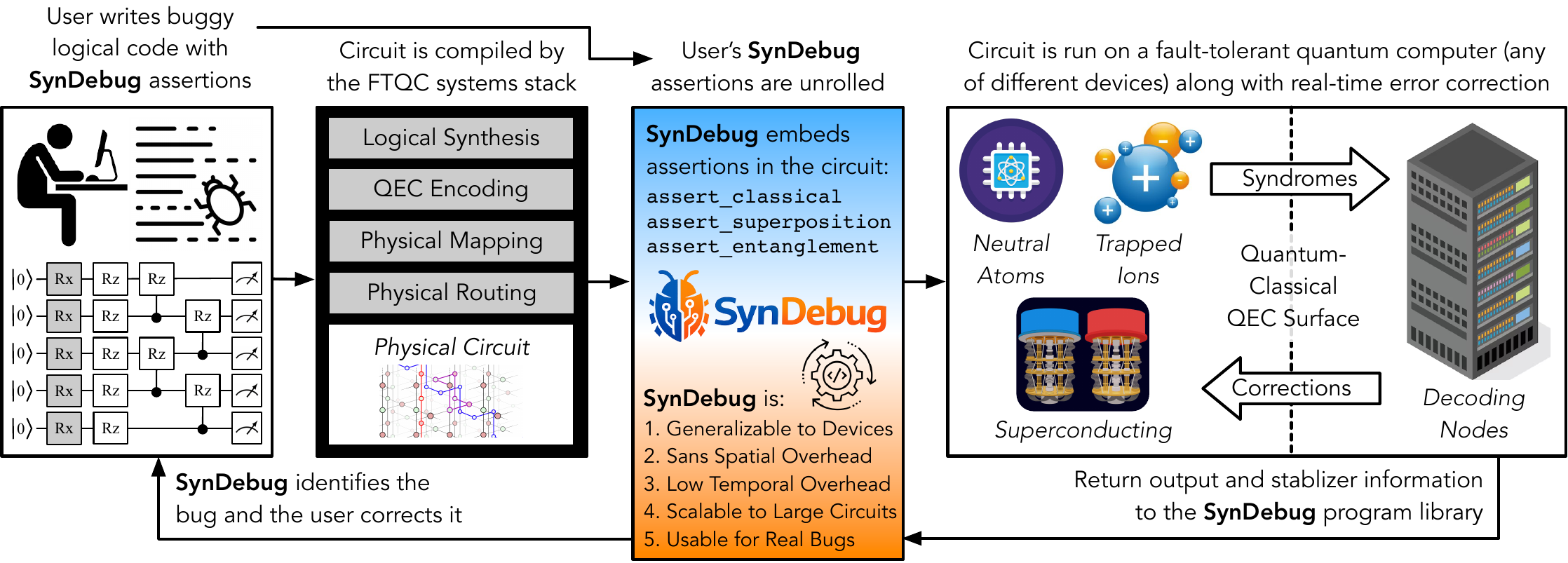}
    \vspace{-3mm}
    \caption{\sol{} embeds lightweight debugging assertions directly into FTQC execution flows while integrating with existing FTQC compilation, QEC, and runtime infrastructures, remaining generalizable across multiple hardware technologies.}
    \label{fig:two_pager}
    \vspace{0mm}
\end{figure*}

\begin{table*}[t]
\centering
\caption{High-level debugging primitives provided by \sol{} and when/how they should be used.}
\label{tab:debugging_methods}
\vspace{-3mm}
\definecolor{headerorange}{RGB}{230,130,0}
\scalebox{0.86}{
\renewcommand{\arraystretch}{1.5}
\begin{tabular}{p{3.5cm}|p{4.8cm}|p{8.0cm}|p{2.6cm}}
\hline
\rowcolor{headerorange!50}
\textbf{Method} & \textbf{What it Does} & \textbf{When to Use} & \textbf{After Assertion} \\
\hline

\texttt{assert\_classical} &
Checks whether a logical qubit register matches an expected computational basis (classical) state. &
Useful for debugging logical state preparation, basis initialization, arithmetic outputs, control-flow conditions, and encoded classical data in quantum registers. &
Execution can continue (eigenstate). \\
\hline
\texttt{assert\_superposition} &
Checks whether a logical qubit register matches an expected superposition (non-classical) state. &
Useful after Hadamard layers, variational ansatz blocks, interference operations, and amplitude-sensitive transformations, where superposition integrity is important. &
Checkpoint (state is destroyed). \\
\hline
\texttt{assert\_entanglement} &
Checks whether an expected entanglement relationship exists between logical qubit registers. &
Useful for debugging Bell-pair generation, Greenberger-Horne-Zeilinger (GHZ) state preparation, encoded communication structures, teleportation routines, and entanglement-based subroutines. &
Checkpoint (state is destroyed). \\
\hline
\end{tabular}}
\vspace{-2mm}
\end{table*}

\textbf{Why it is Challenging to Debug Quantum Computing Programs.} Debugging quantum programs is fundamentally more difficult than debugging classical software because observing program state directly alters the computation itself. In classical systems, developers routinely rely on tracing, memory inspection, print statements, breakpoints, and runtime monitoring to localize logic bugs. In quantum computing, however, measuring a quantum state collapses the underlying superposition, thereby destroying part of the computation \cite{patel2020ureqa,swierkowska2026eccentric}. Also, quantum measurements reveal only partial information about the underlying state, and the resulting outputs are inherently probabilistic. Consequently, determining whether an incorrect output is due to an actual program bug or to expected stochastic behavior is difficult.

Many debugging strategies also rely on simulation, but exact quantum simulation scales exponentially with the number of qubits \cite{tanji2026quantum,patel2026selective}. While this may remain tractable for very small logical circuits, realistic quantum workloads rapidly exceed the capabilities of full-state simulation. These challenges have motivated prior work on runtime assertions, statistical assertions, and lightweight debugging techniques for logical-level NISQ circuits~\cite{li2022exploiting,li2020projection,liu2020quantum,liu2021systematic,kole2026assertion,veganzones2025assertsq,oldfield2025bloch,li2025statistical,huang2019statistical,rovara2025framework2,rovara2025automatically}. Such approaches provide important foundations for quantum software reliability and demonstrate that assertion-based debugging can help identify common logic bugs, including incorrect gate ordering, swapped control-target relationships, missing operations, duplicated gates, and others.

\vspace{2mm}

\noindent\textbf{Why it is More Challenging to Debug Fault-tolerant Quantum Computing Programs.} In FTQC systems, a logical program must be translated into a physical implementation that involves QEC encoding, transversal gates, lattice surgery, magic-state handling, syndrome extraction, decoding, and correction \cite{turner2026scalable,marton2025lattice}. In particular, non-Clifford operations such as the $T$ gate often require expensive auxiliary infrastructures known as magic-state factories, which can dominate both runtime and hardware overhead in practical FTQC systems \cite{sales2025experimental}. Consequently, realistic FTQC execution involves substantial orchestration across quantum hardware, classical decoding systems, patch-level scheduling, and encoded communication structures.

These complexities are especially important in debugging because the executed FTQC implementation can differ dramatically from the user's original logical program. As illustrated in Figure~\ref{fig:bell_pair}, even a simple Bell pair circuit expands into a substantially larger physical implementation involving encoded operations, stabilizer checks, ancilla interactions, and QEC rounds. Debugging at the FTQC level, therefore, requires reasoning simultaneously about logical program semantics and the underlying encoded execution behavior.

Continuing with Bell pair generation as our example, there are several ways in which programming errors can arise, but we will focus on two main types. The first of these might be individual qubit preparation errors. As is the case in many quantum programs, qubits are typically initialized to Z or X basis states; in this instance, the first qubit is initialized to an X basis state (using a Hadamard gate), while the second qubit is initialized to a Z basis state. It is important that this initialization is correct, so that the entangling operation that follows produces the correct Bell pair. The second class of error could be within the entangling operation itself; failure to properly entangle the two qubits will also prevent Bell pair generation, an example of this case is given in Algorithm~\ref{code:example}. These bugs could also be introduced during the compilation from logical to physical circuit. They are easy to identify in this illustrative example; however, debugging quickly becomes untenable for larger circuits. 

Despite the importance of these challenges, existing debugging frameworks do not directly support efficient, effective debugging within FTQC execution flows. \sol{} takes the first step toward targeting this problem (Fig.~\ref{fig:two_pager}).
\section{\sol{}'s High-level Design}
\label{sec:design}

\begin{figure*}
    \centering
    \includegraphics[width=0.99\linewidth]{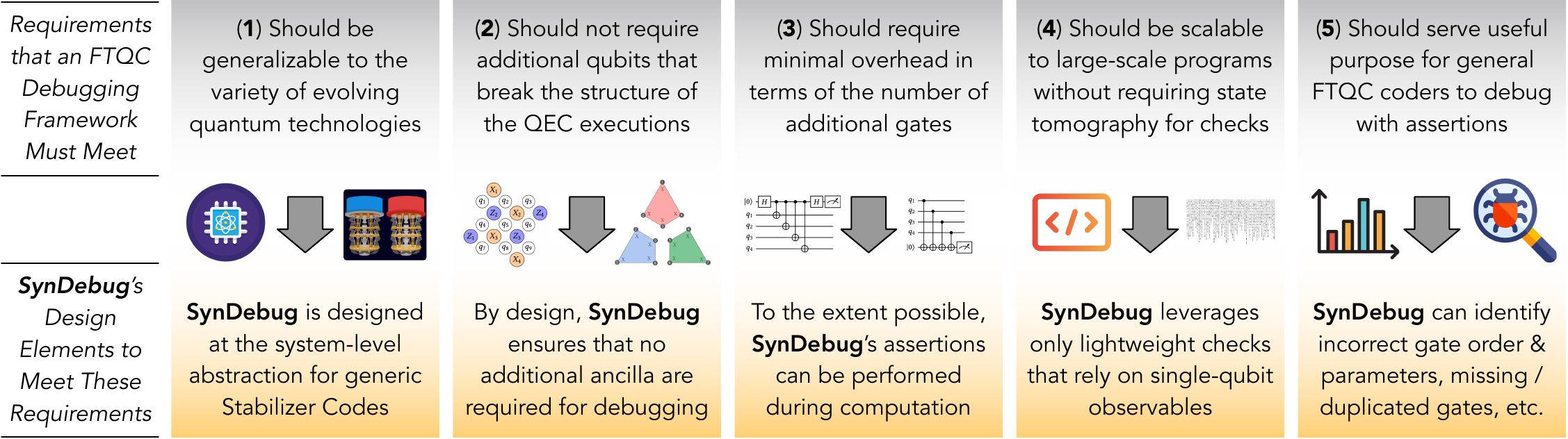}
    \vspace{-3mm}
    \caption{Overview of the requirements that an FTQC debugging framework should meet and how \sol{} satisfies them.}
    \label{fig:design}
    \vspace{-1mm}
\end{figure*}

Figure~\ref{fig:design} provides an overview of the design components of \sol{}.
Guided by practical and scalable requirements, \sol{} introduces three lightweight assertion primitives to debug fault-tolerant programs, (1) \texttt{assert\_classical}, (2) \texttt{assert\_ superposition}, and (3) \texttt{assert\_entanglement}, for validating logical state preparation, superposition structure, and entanglement relationships within FTQC executions. The framework is designed at the stabilizer-code abstraction level to remain compatible with diverse QEC methodologies while avoiding additional qubit overhead and minimizing the execution cost of debugging.

\subsection{Requirements for \sol{}'s Design}
\label{sec:reqs}

Before we describe \sol{}'s design components, we first outline the requirements that we targeted for \sol{}:

\begin{enumerate}[label=\protect\blackcircle{R\arabic*},leftmargin=*]

    \item \textbf{Generalizability Across Hardware+QEC Paradigms.} A debugging framework for FTQC should remain compatible with the rapidly evolving landscape of quantum hardware technologies and QEC methodologies because both quantum devices and QEC implementations continue to evolve substantially across superconducting, neutral-atom, and trapped-ion architectures with differing operational constraints and code structures \cite{zhu2025quantum}.
    
    \item \textbf{Zero Spatial Overhead.} A debugging framework for FTQC should avoid increasing the required qubit footprint because practical FTQC systems already have large physical qubit overheads for logical encoding. Further, hardware is specifically designed and architected for specific QEC patch structures (e.g., IQM's Garnet Chip is designed for the distance-3 surface code shown in Figure~\ref{fig:d3_patch} \cite{krzyszkowski2025analysis} and Google's Willow chip is designed for the distance-7 surface code \cite{google2025quantum}). Thus, additional qubits at locations required by the debugging framework may not physically exist on the hardware.
    
    \item  \textbf{Minimal Temporal Overhead.} A debugging framework for FTQC should introduce as few additional gates, measurements, and execution layers as possible because additional operations increase syndrome extraction requirements and overall circuit depth/runtime.

    \item \textbf{Scalability to Large-Scale FTQC Systems.} A debugging framework for FTQC should scale to large encoded systems without relying on infeasible validation mechanisms because techniques such as full-state or partial-state tomography and exact quantum simulation scale exponentially with system size \cite{tanji2026quantum,patel2026selective}.

    \item  \textbf{Practicality and Developer Usability.} A debugging framework for FTQC should expose practical debugging primitives that integrate naturally into existing software workflows because debugging tools that are difficult to use or disconnected from practitioner needs are unlikely to see adoption in real FTQC software ecosystems. It should be able to detect logic bugs, whether they are programmer mistakes or compiler bugs.
    
\end{enumerate}

\subsection{Design of \sol{}'s Assertions}
\label{sec:assertions}

% \begin{table}
%     \centering
%     \begin{tabular}{ccc}
%      Method Name & Usecase & \\
%      assert_classical & check that part of or entire register is in Z (or X) basis state & Use whenever state is known to be in basis stae\\
%      assert_superposition & check that part of or entire register is in superposition & Use whenever state is known not to be in basis state, or if status is unknown \\
%      assert_entanglement & check that different qubits are entangled & Use after entangling operations between different qubits \\
%     \end{tabular}
%     \caption{Caption}
%     \label{tab:placeholder}
% \end{table}

Table~\ref{tab:debugging_methods} summarizes the primary debugging primitives provided by \sol{} and their intended use cases. These assertions are based on the ones proposed by Huang et al.~\cite{huang2019statistical} for NISQ circuits. Rather than requiring developers to reason about low-level stabilizers, syndrome patterns, or hardware-specific implementation details, \sol{} provides a small set of high-level assertion abstractions that target common classes of FTQC logic bugs. These assertions allow users to verify classical states, superposition structure, and entanglement relationships, which collectively cover many of the errors encountered during FTQC application development, including incorrect gate ordering, missing operations, duplicated operations, and incorrect parameters and orientations.

%A key design goal of \sol{} is to provide these debugging capabilities while remaining practical for large-scale FTQC systems. Consequently, the assertions are implemented using lightweight, observable-based statistical tests that operate at the logical-program level, which are translated into QEC-aware operations on the underlying encoded implementation. This design enables debugging across multiple stabilizer-code families without requiring additional logical qubits or expensive state reconstruction procedures. 

\sol{} uses a simple one-sided binomial test for the \texttt{assert\_classical} and \texttt{assert\_superposition} assertions. For the first of these, we assign the null hypothesis to the outcome of all measurement samples matching the anticipated bitstring of the register (corresponding to a simple basis state). If the null hypothesis can be rejected, the assertion fails, as a significant number of samples are not in the expected state. The \texttt{assert\_superposition} method is similar, but, as no anticipated value is provided, it uses the most probable basis state for the null hypothesis and inverts the test outcome (since superposition is the opposite of being in a basis state). The final assertion, \texttt{assert\_entanglement}, makes use of a contingency table chi-squared ($\chi^2$) analysis, much the same as that of Huang et al.~\cite{huang2019statistical}. Utilizing statistical tests in this fashion has the added advantage of providing confidence bounds for the assertion. 

\vspace{2mm}

\noindent\textbf{Noise and Detection Rates.} Two sources of noise are relevant. Finite-shot sampling noise is handled within the binomial and $\chi^2$ tests. Physical hardware noise is corrected by the QEC layer on which \sol{} operates; any residual logical error rate is accounted for by lowering the fraction of shots expected to match under the null hypothesis. The false-positive rate is thus determined by the significance threshold, while the false-negative rate and the shot count to detection are determined by statistical power. A particular noise model, therefore, affects the shot budget, but not \sol{}'s API, translation, or spatial footprint.

\vspace{2mm}

\noindent\textbf{Eigenstate Measurements.} \sol{} requires only logical-observable readout, and never tomography, copying, or reconstruction. If the expected state is an eigenstate of the measured observable, an ideal measurement preserves it (a standard property of quantum measurement), so the computation can continue. Whether the measurement is physically destructive depends on the platform's lattice surgery, code-specific operations, or hardware~\cite{marton2025lattice}. Note that any ancilla used internally by a machine's standard logical-measurement service is part of that hardware primitive; \sol{} itself never requests a dedicated debugging patch.

\subsection{How \sol{} Meets Each Requirement}

We now briefly describe how our design components address each requirement outlined in Section~\ref{sec:reqs}.

\begin{enumerate}[label=\protect\blackcircle{R\arabic*},leftmargin=*]

    \item \textbf{Generalizability Across Hardware+QEC Paradigms.} To remain relevant in the rapidly evolving QEC landscape, \sol{} is designed to be compatible with a wide range of QEC codes and hardware platforms. To do this, we focus our debugging framework at the system level, i.e., at the level of stabilizer-code abstractions, ensuring compatibility with most of the current QEC landscape. Additionally, designing for generic stabilizer codes enables broad hardware compatibility, as stabilizer codes are leading candidates for QEC across most hardware platforms, including superconducting, neutral-atom, and trapped-ion systems \cite{swierkowska2026eccentric,zhou2025opportunities}. 
    
    \item \textbf{Zero Spatial Overhead.} By design, \sol{} ensures that no extra qubits are added to the physical circuit.
    
    \item \textbf{Minimal Temporal Overhead.} It is crucial that \sol{} be lightweight in terms of the quantum operations required, as high-overhead checks could significantly impact the timing and resource usage. To enable this, we design the assertions, to the extent possible, to be performed during computation. We evaluate this overhead in Section~\ref{sec:demo1} (it is 0\% for most cases).

    \item \textbf{Scalability to Large-Scale FTQC Systems.} \sol{} relies on lightweight checks that are simple measurements of single-logical-qubit observables, often measured on a small subset of logical qubits. Even for a larger subset, the number of qubit observables remains linear with respect to the size of the subset (measuring some fixed number of physical qubits per logical qubit), and sampling them can occur in parallel (measuring different qubits within the same run).

    \item \textbf{Practicality and Developer Usability.} Many quantum programming errors arise from logic mistakes. These mistakes include reversed gate ordering, incorrect gate operands, control-target swaps, missing gates, incorrect parameters, and duplicated gates. \sol{}'s assertions can be utilized to debug all of these error types.
    
\end{enumerate}

\subsection{Scope and Coverage of \sol{}}

\sol{} focuses on detecting logic-level bugs in FTQC programs through lightweight statistical assertions and observable-based debugging primitives. In particular, \sol{} targets scenarios where a quantum program executes successfully from a systems and hardware perspective, yet produces semantically incorrect outputs due to mistakes in the implemented logic. The framework, therefore, provides assertion mechanisms to validate properties such as classical, superposition, and entanglement states during FTQC execution.

At the same time, \sol{} does not attempt to solve several orthogonal or still-open challenges in quantum software reliability. First, \sol{} is not a formal verification framework and does not attempt to mathematically prove correctness properties of arbitrary FTQC programs; instead, it complements existing formal verification and symbolic reasoning approaches~\cite{choudhury2026verification,fang2024symbolic,huang2025efficient}. Second, we do not claim to detect every possible class of quantum software bug, as new types of bugs may arise as quantum computing evolves; however, the primitives introduced by \sol{} will remain useful, as they tackle fundamental software bugs.

Third, \sol{} does not directly target low-level hardware noise or physical fault mechanisms, as these are addressed by the underlying QEC infrastructure. Fourth, \sol{} does not guarantee the elimination of residual logical errors that may persist despite fault-tolerant execution, as reliable large-scale logical error suppression remains an active research challenge \cite{chen2021exponential,putterman2025preserving,google2025quantum}. Finally, \sol{} does not target syntax errors, compile-time crashes, or conventional runtime exceptions, which are typically handled by programming languages, compilers, and classical execution environments before FTQC execution begins \cite{melinte2026runtimespeq,pontolillo2025qucheck,quetschlich2025experience}.

\subsection{Applicability, Generalization, and Discussion}
\label{sec:applicability}

\noindent\textbf{When does \sol{} apply, and how does it generalize?} \sol{} applies at any program point where the developer can state one of three invariants over logical registers: (1) the register is in a known $Z$- or $X$-basis product state, (2) the register is not in a basis state, or (3) two registers are correlated. Such points occur in most quantum programs: after initialization, after arithmetic or oracle blocks driven with basis-state inputs, after ancilla uncomputation, and after entangling steps. For other algorithms, the pattern of Section~\ref{sec:demo3} carries over: test a subroutine with basis-state inputs, and then test the correlations that it creates with its control register. Complex assertions are built by chaining the primitives; e.g., a gate with many control qubits can be defined recursively in terms of smaller controlled gates and debugged piecemeal with smaller \texttt{assert\_entanglement} checks. For other QEC codes and logical-gate realizations, the design requires only the logical Pauli operators of a stabilizer code, and additional NISQ-style assertions can be ported by design. In contrast, \sol{} cannot validate properties that depend on arbitrary amplitudes or relative phases, nor can it detect a bug that preserves every asserted invariant. \textbf{Developer Effort.} Developers reason entirely at the logical level: each assertion is a single API call, and \sol{} performs the physical-qubit translation, measurement insertion, statistical testing, and result reporting. We note that we did not conduct a formal human study with scientists and developers regarding the applicability of \sol{}. We leave this as future work. \textbf{Bug Corpus.} \sol{} targets logic bugs whose classes already exist at the logical (NISQ) level; QEC encoding complicates their physical manifestation but does not introduce a new bug taxonomy. A separate FTQC-specific corpus would therefore duplicate established resources~\cite{zhao2023bugs4q,campos2021qbugs,paltenghi2022bugs} and would require a compiler-specific corpus for each code family. \textbf{Native Debugging Support.} Native FTQC debugging support is attractive but architecturally expensive, because dedicated measurement, control, and debugging resources must be provisioned across the hardware, the QEC schedule, and the classical control system. \sol{} does not depend on such changes; nonetheless, if future hardware provides native debugging primitives, the backend can map \sol{}'s assertions onto them without changing its programming model.

\section{Implementation Details of \sol{}}
\label{sec:implementation}

\begin{figure*}[t]
    \centering
    \includegraphics[width=0.99\linewidth]{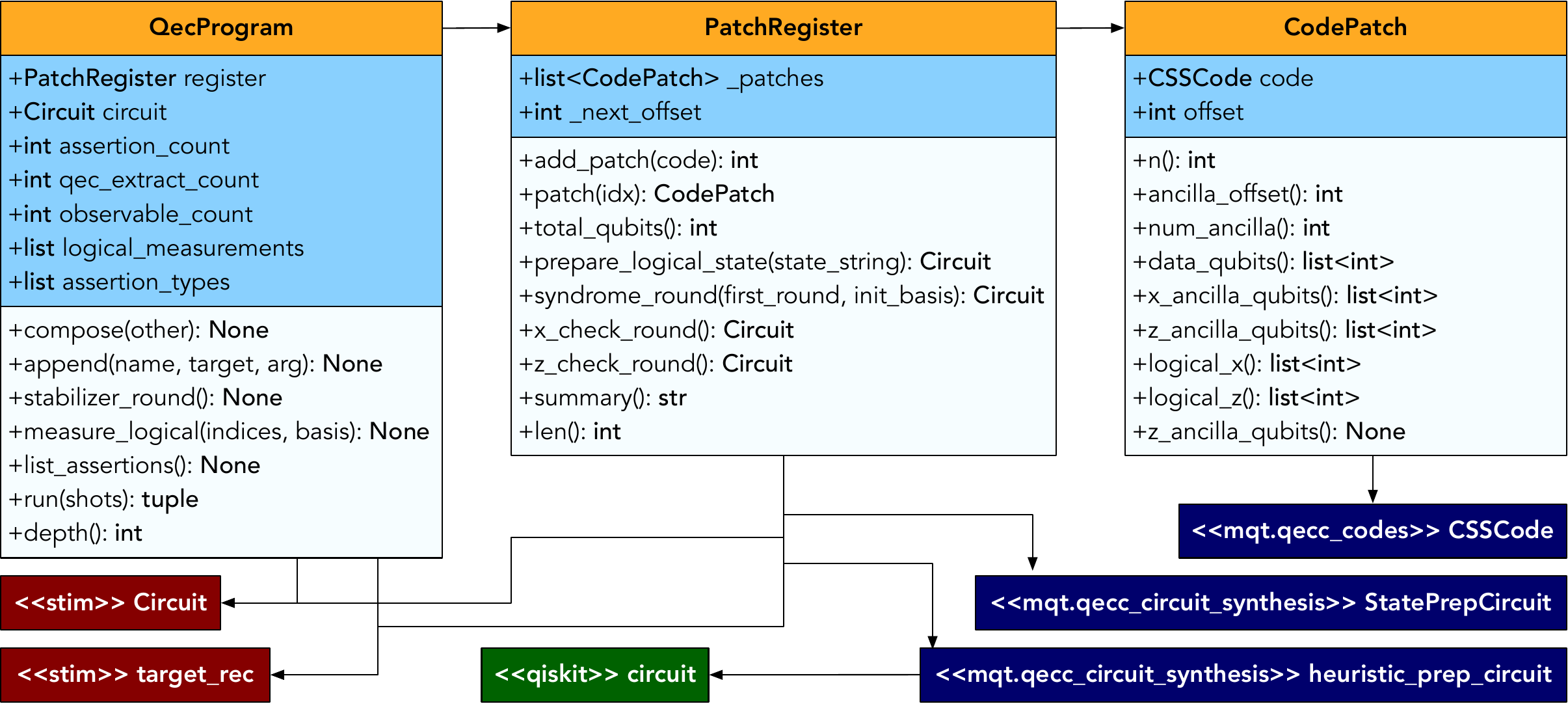}
    \vspace{-3mm}
    \caption{System architecture of the \sol{} debugging framework with three new classes and external-library integration.}
    \label{fig:details}
    \vspace{-1mm}
\end{figure*}

Figure~\ref{fig:details} provides an overview of the implementation architecture and class structure of \sol{}.

\vspace{2mm}

\noindent\textbf{Class Structure.} The module defines a four-level abstraction hierarchy for managing multi-patch CSS code circuits, layered atop the MQT QECC library's code-theoretic primitives. Each level encapsulates a distinct concern (i.e., physical qubit allocation, single-patch addressing, register-wide composition, and program-level orchestration) and the boundaries between them are made explicit through a consistent set of indexing conventions that permit simultaneous addressability at both the physical and logical levels.

\vspace{2mm}

\noindent\textbf{Foundational Dependencies.} The construction rests on three external primitives. The \texttt{CSSCode} class from \texttt{mqt.qecc. codes} supplies the algebraic data of a code: the parity-check matrices \texttt{Hx} and \texttt{Hz}, a distance parameter, and methods returning the logical Pauli operators as strings over the data qubits. The \texttt{heuristic\_prep\_circuit} routine and its associated \texttt{StatePrepCircuit} wrapper produce non-fault-tolerant logical state-preparation circuits as Qiskit \texttt{QuantumCircuit} objects, which are subsequently transpiled to Stim format via the local \texttt{qiskit\_to\_stim\_circuit} helper (adapted from MQT QECC). The module thus consumes code definitions and preparation circuits from MQT QECC while producing executable Stim circuits suitable for downstream simulation or noise analysis. We remark that we restrict our implementation to codes having ($k = 1$). This restriction allows us to greatly simplify the indexing of our logical program, as there is a one-to-one mapping between logical qubits and error correction patches: each patch corresponds to a single, unambiguous logical qubit.

\vspace{2mm}

\noindent\textbf{\texttt{CodePatch}: Single-Logical Qubit Addressing.} \texttt{CodePatch} is the lowest abstraction layer above the bare \texttt{CSSCode}. It is a frozen-by-convention dataclass associating a code instance with a single integer, \texttt{offset}, denoting the index of its first data qubit within an enclosing global register. From this single offset, the class derives a complete physical addressing scheme through a small set of properties and methods. The indexing convention within a patch is strictly contiguous and ordered: the $n$ data qubits occupy indices $[\texttt{offset},\, \texttt{offset} + n)$, immediately followed by the $X$-stabilizer ancillas at $[\texttt{ancilla\_offset},\, \texttt{ancilla\_offset} + \texttt{Hx.shape[0]})$, and then the $Z$-stabilizer ancillas at the subsequent block. Consequently, given only a patch and a stabilizer row index, the corresponding physical ancilla qubit is determined arithmetically without any auxiliary lookup.

Logical-level addressing is exposed through \texttt{logical\_x()} and \texttt{logical\_z()}. These methods translate the symbolic Pauli strings returned by \texttt{CSSCode.x\_logicals\_as\_pauli\_ strings()} and \texttt{CSSCode.z\_logicals\_as\_pauli\_strings()} into concrete physical qubit indices, by enumerating the support of the logical operator within the data block and offsetting it into the global register. This is the key mechanism by which logical operators (i.e., abstract objects defined by the QEC code) become directly applicable as gate targets in a circuit acting on physical qubits.

\vspace{2mm}

\noindent\textbf{\texttt{PatchRegister}: Multi-Patch Composition.} \texttt{PatchRegister} aggregates an ordered list of \texttt{CodePatch} instances and manages the allocation of contiguous physical qubit ranges across them. The \texttt{\_next\_offset} field functions as a bump allocator: each call to \texttt{add\_patch} records the current offset as the new patch's base, then advances the offset by the patch's full footprint (data plus all ancillas). This guarantees that no two patches overlap, that the global qubit count is always \texttt{total\_qubits()}, and that physical indices remain stable for the lifetime of the register.

The register's preparation method, \texttt{prepare\_logical\_ state}, demonstrates how the layered indexing pays off. It accepts a character-per-patch state string (\texttt{'0'}, \texttt{'1'}, \texttt{'p'}, \texttt{'m'}) (corresponding to the states $\ket{0}, \ket{1}, \ket{+}, \ket{-}$ respectively) and constructs a Qiskit circuit acting on the full register. For each patch, it invokes \texttt{heuristic\_prep\_circuit} to obtain a local preparation routine, then composes that routine into the global circuit at the qubit range $[\texttt{offset},\, \texttt{offset} + n)$. Logical-basis flips required for $\ket{1}_L$ and $\ket{-}_L$ are implemented by applying physical $X$ or $Z$ gates to the support of the logical operator, as obtained from the patch's \texttt{logical\_x()} or \texttt{logical\_z()} method.

The result is converted to a Stim circuit and returned. Notably, the same method makes use of both indexing levels in a single body: (1) physical ranges for circuit composition, and (2) logical supports for basis flips.

\vspace{2mm}

\noindent\textbf{\texttt{QecProgram}: Error Corrected Program-Level.} \texttt{QecProgram} is a thin wrapper over \texttt{PatchRegister} that captures the common case of a homogeneous register: \texttt{n\_patches} copies of a single code, prepared in a specified joint logical state. It exposes two fields to the user: (1) the underlying \texttt{register} and (2) the prepared \texttt{circuit}, and provides a \texttt{compose} operator for appending further Stim circuit fragments (such as logical gate gadgets or syndrome-extraction rounds) onto the prepared state.

This layer is intentionally minimal; it does not introduce new addressing conventions but rather bundles the register and its initial circuit into a single compilation unit.

\vspace{2mm}

\noindent\textbf{Indexing Conventions and Dual Addressability.} The design introduces two coexisting addressing modes that reference the same underlying physical register:

At the \emph{physical level}, every qubit in every patch has a unique global integer index, determined by the patch's offset and the within-patch convention (data, then $X$-ancilla, then $Z$-ancilla). QEC codes that need to emit physical gates, such as stabilizer extraction circuits, transversal operations, and measurement schedules, obtain these indices via \texttt{data\_ qubits()}, \texttt{x\_ancilla\_qubits()}, and \texttt{z\_ancilla\_qubits()}.

At the \emph{logical level}, each patch represents a single logical qubit, identified by its index in the register. Logical Pauli operators are obtained by querying the patch and are returned as lists of physical indices already translated into the global register's coordinate system. This permits logical-level reasoning (``apply $X_L$ to patch 2'') to be expressed at the call site, while the patch object handles the translation to physical gate targets in keeping with FTQC constraints.

The strength of this two-level scheme is that neither mode requires an auxiliary mapping table. The physical indexing is purely arithmetic in the offsets, and the logical indexing is derived on demand from the \texttt{CSSCode}'s own representation of its logicals. The same \texttt{CodePatch} object simultaneously serves as a physical-layout descriptor and as a logical-qubit handle, with the offset acting as the single point of coupling between the two views.

\vspace{2mm}

\noindent\textbf{Initialization Flow.} The initialization of a multi-patch logical state proceeds through a fixed three-stage pipeline, with each stage operating at a distinct level of the abstraction hierarchy. The pipeline is invoked end-to-end by the \texttt{QecProgram} constructor, but its stages are independently meaningful and may be driven directly through the \texttt{PatchRegister} interface when finer control is required.

\vspace{2mm}

\noindent\textbf{Stage 1: Register Population.} The first stage builds the physical layout. A fresh \texttt{PatchRegister} is instantiated with an empty patch list and \texttt{\_next\_offset} initialized to zero. For each code instance to be included, \texttt{add\_patch} is invoked: the method constructs a \texttt{CodePatch} with the current value of \texttt{\_next\_offset} as its base, appends it to the register, and advances \texttt{\_next\_offset} by the patch's full footprint $n + \texttt{Hx.shape[0]} + \texttt{Hz.shape[0]}$. After all patches are added, \texttt{total\_qubits()} reports the global register size, and every patch holds a stable, non-overlapping physical address range. No quantum circuit has yet been constructed; this stage is purely an allocation pass.

\vspace{2mm}

\noindent\textbf{Stage 2: Per-patch State Preparation.} The second stage is driven by \texttt{prepare\_logical\_state}, which constructs a Qiskit \texttt{QuantumCircuit} over \texttt{total\_qubits()} qubits and iterates through the patches in register order, consuming one character of the state string per patch. For each character, the method selects two booleans: \texttt{zero\_state}, which determines whether the underlying preparation circuit targets the computational basis ($\ket{0}_L$, $\ket{1}_L$) or the Hadamard basis ($\ket{+}_L$, $\ket{-}_L$), and \texttt{flip}, which indicates whether a subsequent logical-basis flip is required. The MQT QECC routine \texttt{heuristic\_prep\_circuit} is then invoked on the patch's code with the chosen \texttt{zero\_state} value, returning a non-fault-tolerant preparation circuit on $n$ qubits. This local circuit is composed into the global circuit at the patch's data-qubit range $[\texttt{offset},\, \texttt{offset} + n)$ via Qiskit's \texttt{compose}. If \texttt{flip} is set, the appropriate logical operator support is fetched (\texttt{logical\_x()} for computational-basis flips, \texttt{logical\_z()} for Hadamard-basis flips) and a physical $X$ or $Z$ gate is applied to each qubit in that support. After the loop, the assembled Qiskit circuit is finally passed through the conversion method \texttt{qiskit\_to\_stim\_circuit}, which performs a gate-by-gate translation into a Stim \texttt{Circuit}.

\vspace{2mm}

\noindent\textbf{Stage 3: Program Assembly.} When initialization is driven through \texttt{QecProgram}, the third stage simply binds the populated register and the prepared Stim circuit as fields of the program object, and exposes the \texttt{compose} method for appending subsequent circuit fragments. 

Direct users of \texttt{PatchRegister} skip this stage and receive the Stim circuit as a return value.

\vspace{2mm}

\noindent\textbf{Properties of the Initialization Pipeline.} Three features of this flow are worth noting. First, the separation of allocation (Stage 1) from circuit construction (Stage 2) means that the physical addressing scheme is fully determined before any gate is emitted, so the same \texttt{PatchRegister} can be reused to construct multiple distinct initial circuits without re-allocating qubit indices. Second, the use of \texttt{heuristic\_ prep\_circuit} makes the initialization non-fault-tolerant by construction; the pipeline is intended as the starting point for circuits that will subsequently apply syndrome-extraction rounds rather than as a verified preparation gadget. Third, the per-patch independence of Stage 2 (each patch's preparation circuit acts only on its own data qubits, and basis flips touch only the logical operator support of that patch) means that the initial multi-patch state is a tensor product of single-patch logical states, with no inter-patch entanglement introduced at initialization. Any joint logical state required by the program must be produced by circuit fragments composed onto the program after initialization.

\vspace{2mm}

\noindent\textbf{User-facing APIs and Instrumentation.} \sol{} exposes a small set of debugging primitives that integrate directly into FTQC programs. The primary user-facing APIs are \texttt{assert\_classical}, \texttt{assert\_superposition}, and \texttt{assert\_ entanglement}, which enable developers to verify logical basis states, superposition structure, and entanglement relationships, respectively. These assertions operate at the logical-program level while being automatically translated into QEC-aware checks over the underlying encoded implementation. Additional helper APIs support assertion insertion, result collection, and integration with existing FTQC toolchains such as Stim and MQT.

In addition to providing debugging assertions, \sol{} includes lightweight instrumentation and monitoring capabilities that track assertion usage throughout program development, and these records are pushed up to the \texttt{QecProgram} wrapper level. Specifically, the framework records and reports to the user the total number of assertions inserted into a program, the number of assertions of each type (e.g., \texttt{assert\_entanglement}, \texttt{assert\_classical}), and the associated overheads introduced by debugging. This information allows developers to identify heavily instrumented code regions, quantify debugging costs, and progressively remove assertions as bugs are addressed.
\section{Case Studies of \sol{}'s Usage}
\label{sec:demos}

\begin{algorithm}[t]
\caption{Initialization with \sol{} assertions.}
\label{code:init}
\vspace{-2mm}
\begin{tcolorbox}[boxrule=0.2mm, colframe=black, colback=blue!5!white]
\begin{lstlisting}[style=quantumcode]
# Initialize program with three surface code patches
# Initialize register as integer 2 
code = CSSCode.from_code_name("Surface", 3)
prog = QecProgram(code, (*@3@*), "010") 
# Add syndrome extraction round
prog.stabilizer_round()
# Assertions verifying initializations
assert_classical(prog, value=(*@2@*))
# Quantum gate operations using initialized register...
\end{lstlisting}
\end{tcolorbox}
\vspace{-1mm}
\end{algorithm}

We now present three case studies that demonstrate how \sol{} can be used. We perform our experiments using classical simulations of quantum programs, as FTQC hardware does not yet exist. All experiments were run on an Apple MacBook Pro equipped with the M1 Pro CPU and 16 GB of memory, running macOS 26.5. All methods were implemented in Python 3.12.1. Circuits and simulations were created and performed using Stim~\cite{stim}, a library for high-performance stabilizer simulation. Codeword preparation is performed using the Munich Quantum Toolkit QECC module~\cite{mqt,mqtstateprep}, which outputs circuits in Qiskit~\cite{qiskit} format. 

\subsection{Demonstration 1: State Preparation Debugging}
\label{sec:demo1}

\noindent\textbf{Methodology.} Our first case study concerns the preparation of code patches on a computer. This preparation stage involves generating a multi-qubit state on the code's physical qubits. This multi-qubit state corresponds to a known logical state (codeword), which is typically a basis state $\ket{0}$ or $\ket{1}$. 

Initializing a register of error-corrected qubits entails running codeword preparation on each code patch, resulting in a simple product state of the logical qubits. Our first case study demonstrates the utility of assertion checking in catching program bugs during the initialization phase.

Typically, the initial state of a program is a product state of simple single-qubit states. This simple structure allows us to easily verify logical initialization by performing computational-basis measurements. To do this, we make use of \texttt{assert\_ classical}, which checks that the state matches the binary representation of the specified integer. We give an example of this in Algorithm~\ref{code:init}, which shows the initialization of multiple surface code patches. The patches are initialized so that the logical register is in the $\ket{010}$ state, representing the integer 2. On line 8, we have the assertion, which performs logical measurements using data qubits. The assertion is then followed by logical operations using the initialized register (the code can continue if the assertion is correct). 

Generally, as the correct initial state is an eigenstate of the measured observable, measuring the observable will not destroy any quantum superposition or information in the system (assuming a non-destructive measurement apparatus). This means the computation can continue without resetting, and the test statistics can be computed during post-processing. Of course, it is also possible to check separate blocks of the program piece by piece by removing the parts that follow the assertion. This flexibility in approach allows the user to prioritize overall shot count/budget or the runtime of individual shots as needed.

We also note that it is sometimes useful for the user to initialize in the Hadamard basis of $\{\ket{+},\, \ket{-} \}$. To accommodate this, we also allow the user to specify the basis used during the assertion check. The process is the same as in the $Z$-basis, except that the measurements are converted to $ X$-basis measurements. To this end, the method allows the user to specify a ``state string'' of the form $s \in \{0, 1, p, m\}^L$, where $L$ is the number of logical qubits and $0, 1, p, m$ denote the states $\ket{0}, \ket{1}, \ket{+}, \ket{-}$ respectively, allowing the user to initialize the register to a product state of $Z$ and $X$ basis states. Initialization assertions can also be performed using \texttt{assert\_superposition}, to check that qubits within the register are in a superposition. However, doing so is inherently worse than the \texttt{assert\_classical} case, as measuring a quantum state in an unknown superposition destroys the superposition, rendering it unusable in subsequent calculations. We therefore recommend judicious use of \texttt{assert\_superposition} only when necessary.
Note: here is flexibility in the choice of assertion: the $\ket{+++}$ control register of Shor's algorithm (Section~\ref{sec:demo3}) could be checked with \texttt{assert\_superposition}, but \texttt{assert\_classical} in the $X$ basis is stronger because it verifies the intended $\ket{+}$ eigenstate and allows the program to continue.

\begin{table}[t]
\centering
\caption{Circuit depths with and without \sol{} for some examples of classical states for one round of QEC. \sol{} maintains a low debugging overhead (1.8\% on average), with the overhead footprint decreasing with more rounds. The reported depths also hold for \texttt{assert\_superposition}, which performs the same measurements.}
% CAMERA-READY NOTE (authors): the listed overheads are relative to the depth WITH SynDebug (e.g., Surface d=3 p0: 2/14 = 14.3%).
% Relative to the uninstrumented depth they are 2/12 = 16.7%, 1/36 = 2.8% (Golay), and 1/14 = 7.1% (Hex. Color). Consider recomputing this column and the 1.8% average.
\label{tab:classical_init}
\vspace{-3mm}
\definecolor{headerblue}{RGB}{40,100,230}
\definecolor{headerorange}{RGB}{230,130,0}
\scalebox{0.86}{
\begin{tabular}{|lll|p{15mm}|p{15mm}|p{15mm}|}%{|lll|p{14mm}|p{14mm}|p{14mm}|p{14mm}|p{14mm}|p{14mm}|}
\hline
\rowcolor{headerorange!50}
 &  &  & \multicolumn{3}{c|}{\textbf{Circuit Depth}}  \\
\rowcolor{headerorange!50}
\textbf{QEC Code} & \textbf{$d$} & \textbf{State} & Without \sol & With \ \ \ \sol & Percent Overhead \\
\hline
Steane & 3 & 0 & 14 & 14 & 0.0 \\
Steane & 3 & 10 & 15 & 15 & 0.0 \\
Shor & 3 & 1 & 14 & 14 & 0.0 \\
Shor & 3 & 100 & 14 & 14 & 0.0 \\
Tetrahedral & 3 & 0 & 19 & 19 & 0.0 \\
Golay & 7 & m & 36 & 37 & 2.7 \\
Surface & 3 & 0 & 12 & 12 & 0.0 \\
Surface & 3 & p0 & 12 & 14 & 14.3 \\
Surface & 5 & 0 & 13 & 13 & 0.0 \\
Surface & 5 & pm & 16 & 16 & 0.0 \\
Hex. Color & 3 & 0 & 14 & 14 & 0.0 \\
Hex. Color & 5 & 1 & 19 & 19 & 0.0 \\
Hex. Color & 7 & 0 & 20 & 20 & 0.0 \\
Hex. Color & 3 & 0p1 & 14 & 15 & 6.7 \\
Sq. Oct. Color & 3 & 0 & 13 & 13 & 0.0 \\
Sq. Oct. Color & 5 & p & 18 & 18 & 0.0 \\
Sq. Oct. Color & 3 & pm & 14 & 14 & 0.0 \\
\hline
\rowcolor{headerblue!50}\multicolumn{5}{l|}{\textbf{Average} (for all states inc. ones not listed above)} & \textbf{1.8\%} \\
\hline
\end{tabular}}
\vspace{-1mm}
\end{table}

\vspace{2mm}

\noindent\textbf{Results.} Table~\ref{tab:classical_init} quantifies the circuit-depth overhead introduced by \sol{} during logical state initialization. Across all evaluated QEC codes, code distances, and target states (only some randomly-selected samples listed here), \sol{} incurs only a modest average overhead of 1.8\%, with many configurations exhibiting no measurable increase in circuit depth. These results demonstrate that initialization assertions can be incorporated into FTQC programs without substantially increasing the cost of encoded execution. We note that these overhead figures are relative to one QEC round (we selected one round to show the worst-case overhead); as the number of rounds increases, the relative overhead would become negligible.

\begin{figure}[t]
    \centering
    \includegraphics[width=0.99\linewidth]{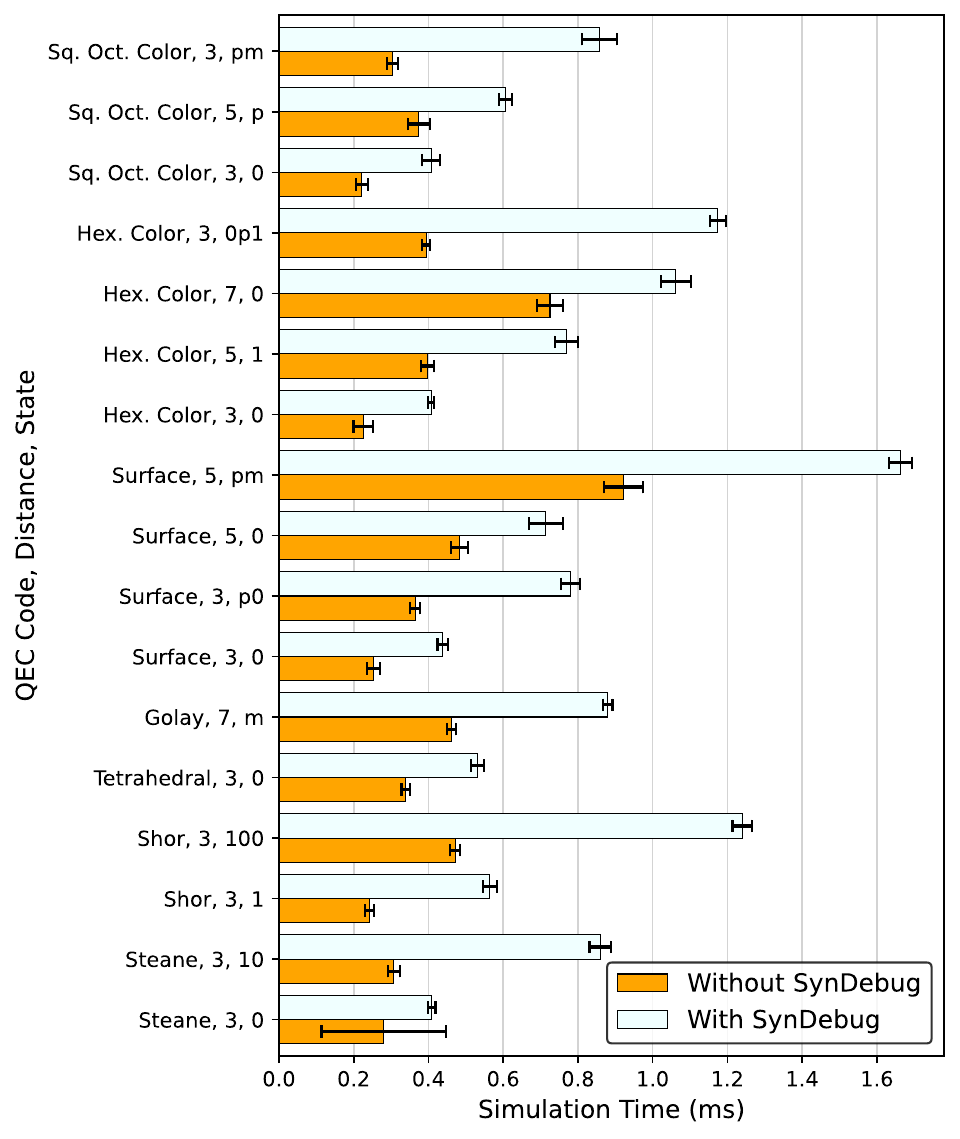}
    \vspace{-5mm}
    \caption{While \sol{} increases the simulation times, they are sufficiently small (milliseconds) to perform FTQC debugging even on simulation targets.}
    \label{fig:sim_time}
    \vspace{-1mm}
\end{figure}

Figure~\ref{fig:sim_time} reports the corresponding simulation times of these circuits. We note that FTQC hardware does not yet exist; therefore, we simulate the circuits on classical computers. Although \sol{} increases simulation runtime, the absolute overhead remains small, requiring only milliseconds per execution across all evaluated configurations. This result is important because simulation often serves as the first line of debugging before deployment on quantum hardware. Developers may prefer simulation when hardware access is limited, queue times are long, execution costs are high, or rapid debugging iterations are needed during development. The low simulation overhead, therefore, enables practical assertion-based debugging workflows while preserving the fast turnaround times necessary for effective software development. We also note that we do not anticipate the real-hardware runtime overhead to be proportional to the simulation overhead; in fact, we expect it to be minuscule because real hardware supports \sol{}'s checks natively for parallel execution, while the Stim-based simulation requires additional checks that make it less efficient than the default stabilizer-based Stim implementation.

Lastly, we note that while directly measuring the impact of \sol{} on the logical error rate (LER) is challenging because many assertions naturally coincide with measurements already performed by the target program, we note that any increase in LER is upper-bounded by the additional circuit operations introduced by \sol{}. Since \sol{} incurs only a small circuit-depth overhead (1.8\% on average and 0\% for most cases), any corresponding increase in logical failure probability is expected to be negligible. Note: we do not report further metrics for brevity as they are derivatives of the reported ones or are unaffected by \sol{}: the number of additional debugging qubits is zero, the gate-count and QEC-round overheads follow from the added circuit depth, and the detection rates are determined by the statistical tests.

\begin{algorithm}[t]
\caption{Entangling qubits with \sol{} assertions.}
\label{code:ent}
\vspace{-2mm}
\begin{tcolorbox}[boxrule=0.2mm, colframe=black, colback=blue!5!white]
\begin{lstlisting}[style=quantumcode]
# Initialize program with two surface code patches
# Initialize register as |+>|0>, so that output
# is Bell state |00> + |11>
code = CSSCode.from_code_name("Surface", 7)
prog = QecProgram(code, 2, "p0") 
# Add syndrome extraction round
prog.stabilizer_round()
# Add CNOT gadget from logical 0 to 1
transversal_cnot(prog, 0, 1)
# Check errors after logical operation
prog.stabilizer_round()
# Assertions verifying entanglement
assert_entanglement(prog, subset_a=[0], subset_b=[1])
# Quantum gate operations using Bell pair...
\end{lstlisting}
\end{tcolorbox}
\vspace{-1mm}
\end{algorithm}

\subsection{Demonstration 2: Entanglement Checking}
\label{sec:demo2}

\noindent\textbf{Methodology.} Our second case study evaluates the overhead of assertion-based checking for entangling operations in the context of error-corrected quantum programs. To construct a comprehensive testing suite, we initialize the target qubit in one of the two of the X-basis states and the control qubit in one of the Z-basis states, yielding the four possible Bell pairs as post-entanglement targets. We further vary the code distance across several values to assess scalability, and for each configuration, we report the resulting circuit depth both with and without the inserted assertion measurements.

The assertion workflow for this case study is illustrated in Algorithm~\ref{code:ent}. We begin by instantiating a quantum error correction program over two surface code patches at a given code distance. The working register is initialized in the state $\ket{+}\ket{0}$, such that applying a logical CNOT from the first patch to the second produces the Bell state $\frac{1}{\sqrt{2}}(\ket{00}+\ket{11})$. Syndrome extraction rounds are inserted before and after the transversal CNOT gadget to enable error detection at the physical level. Following the entangling operation, we invoke \texttt{assert\_entanglement} on the two logical qubits. The assertion outcome is determined statistically. Specifically, we construct a contingency table from the observed qubit outcomes across repeated executions and apply a chi-squared ($\chi^2$) test to determine whether the observed distribution is consistent with the expected correlations of the target Bell state. This statistical formulation naturally accommodates the shot noise inherent in finite-sample quantum measurement and provides a principled threshold for accepting or rejecting the assertion. We note, however, that measuring individual qubits within an entangled system destroys the multiqubit state, so program execution cannot simply resume from the point of measurement. Instead, the assertion point must be treated as a checkpoint, which is then removed after the relevant operations have passed the assertion. A destructive assertion acts as a checkpoint only when execution must proceed beyond that location, and the repeated shots at a checkpoint do not require the developer to manually modify the program for each execution. While this introduces a cost in terms of total execution count, it does not affect the depth or structure of any individual circuit instance, preserving the fidelity characteristics of the original program.

\vspace{2mm}

\begin{table}[t]
\vspace{-2mm}
\centering
\caption{Circuit depths with and without \sol{} for entangled states for one round of QEC. For all entangling circuits trialed, \sol{} adds no overhead in circuit depth.}
% CAMERA-READY NOTE (authors): the row "Hex. Color & 3 & p0" appears twice in this table.
\label{tab:entanglement}
\vspace{-3mm}
\definecolor{headerblue}{RGB}{40,100,230}
\definecolor{headerorange}{RGB}{230,130,0}
\scalebox{0.86}{
\begin{tabular}{|lll|p{15mm}|p{15mm}|p{15mm}|}%{|lll|p{14mm}|p{14mm}|p{14mm}|p{14mm}|p{14mm}|p{14mm}|}
\hline
\rowcolor{headerorange!50}
 &  &  & \multicolumn{3}{c|}{\textbf{Circuit Depth}}  \\
\rowcolor{headerorange!50}
\textbf{QEC Code} & \textbf{$d$} & \textbf{State} & Without \sol & With \ \ \ \sol & Percent Overhead \\
\hline
Steane & 3 & p0 & 24 & 24 & 0.0 \\
Steane & 3 & m0 & 25 & 25 & 0.0 \\
Shor & 3 & p0 & 23 & 23 & 0.0 \\
Shor & 3 & p1 & 24 & 24 & 0.0 \\
Tetrahedral & 3 & p0 & 33 & 33 & 0.0 \\
Golay & 7 & m1 & 55 & 55 & 0.0 \\
Surface & 3 & p0 & 20 & 20 & 0.0 \\
Surface & 3 & m0 & 21 & 21 & 0.0 \\
Surface & 5 & p0 & 23 & 23 & 0.0 \\
Surface & 5 & m1 & 24 & 24 & 0.0 \\
Hex. Color & 3 & p0 & 23 & 23 & 0.0 \\
Hex. Color & 5 & p1 & 29 & 29 & 0.0 \\
Hex. Color & 7 & m0 & 30 & 30 & 0.0 \\
Hex. Color & 3 & p0 & 23 & 23 & 0.0 \\
Sq. Oct. Color & 3 & p0 & 22 & 22 & 0.0 \\
Sq. Oct. Color & 5 & p0 & 30 & 30 & 0.0 \\
Sq. Oct. Color & 3 & p1 & 22 & 22 & 0.0 \\
\hline
\rowcolor{headerblue!50}\multicolumn{5}{l|}{\textbf{Average} (for all states inc. ones not listed above)} & \textbf{0.0\%} \\
\hline
\end{tabular}}
\vspace{-1mm}
\end{table}

\begin{algorithm}[t]
\caption{Shor's algorithm debugging with \sol{}.}
\label{code:shor}
\vspace{-2mm}
\begin{tcolorbox}[boxrule=0.2mm, colframe=black, colback=blue!5!white]
\begin{lstlisting}[style=quantumcode]
N = 15 # number to be factored
a = 7 # base used in arithmetic
# Initialize program with 7 surface code patches. 
# Control register as |+++>
# Working register as integer 1
code = CSSCode.from_code_name("Surface", 3)
prog = QecProgram(code, 7, "ppp1000") 
c_id = [0, 1, 2]
w_id = [3, 4, 5, 6]
# Add syndrome extraction round
prog.stabilizer_round()
# Check initialization of two registers
assert_classical(prog, indices=c_id, value=0, basis="X")
assert_classical(prog, indices=w_id, value=1)
# Check one of the arithmetic ops and entanglement
code = CSSCode.from_code_name("Surface", 3)
prog = QecProgram(code, 7, "0101000") 
# Add syndrome extraction round
prog.stabilizer_round()
# Apply mod mult x -> x*7^2 mod 15
c_mod15(prog, a, power=2)
# Add syndrome extraction round
prog.stabilizer_round()
# First, assert the working register is in the correct
# state, 7^2*1 mod 15 = 4
assert_classical(prog, indices=w_id, value=4)
# Check that the control register is entangled with 
# working register
assert_entanglement(prog, subset_a=c_id, subset_b=w_id)
# Repeat for different powers...
\end{lstlisting}
\end{tcolorbox}
\vspace{-1mm}
\end{algorithm}

\noindent\textbf{Results.} Table~\ref{tab:entanglement} quantifies the circuit-depth overhead introduced by \sol{} for entanglement-based assertions. Across all evaluated QEC codes, code distances, and target states, \sol{} incurs no overhead in terms of circuit depth. This result is significant from a practical standpoint, as it demonstrates that \sol{} does not increase the critical-path length of the compiled program, a key concern in error-corrected settings where deeper circuits compound the risk of logical failure. This is because the measurements required for assertion checking by \sol{} can be performed during the QEC syndrome extraction by design, without incurring any additional depth overhead.

\subsection{Demonstration 3: Shor's Algorithm Debugging}
\label{sec:demo3}

Our third case study applies the assertion framework to Shor's algorithm, a key quantum algorithm that has a proven exponential quantum advantage in integer factorization for cryptographic applications \cite{shor1994algorithms}, demonstrating how the relatively simple assertion primitives developed in the preceding case studies can be composed to support the debugging of substantially more complex quantum programs. This case study illustrates how \sol{}'s assertions can be deployed in concert at multiple program points to build a layered diagnosis of algorithmic correctness and detect logic bugs at the program or compilation level.

The core of Shor's algorithm relies on a sequence of controlled modular multiplication operations acting on a working register that is initialized to a known basis state. We begin by applying \texttt{assert\_classical} to this register immediately after initialization, verifying that the register has been correctly prepared in the expected computational basis state before any arithmetic is performed. Following the application of the controlled modular multiplication operations, we invoke \texttt{assert\_classical} a second time to confirm that the arithmetic units are producing the expected output values. This pair of classical assertions bookends the modular arithmetic stage of the algorithm, providing confidence that the reversible classical logic is functioning correctly in isolation.

Beyond verifying the arithmetic itself, we leverage \texttt{assert\_ entanglement} to validate the relationship between the control and working registers. Because each control qubit governs a distinct modular multiplication corresponding to a different power of the base, each control qubit becomes entangled with a different integer state in the working register after its respective controlled operation. Asserting these entanglement relationships confirms that the control register is properly correlated with the working register prior to the application of the inverse quantum Fourier transform. This is a critical verification point, as the correctness of the final period-finding measurement depends on the phase information encoded in the control register by these entangled correlations. Any fault in the controlled arithmetic or in the entanglement structure would corrupt the phase kickback and, consequently, the algorithm's output.
Taken together, these assertions demonstrate that a small set of composable primitives (classical state checking and entanglement validation) is sufficient to instrument a program of non-trivial algorithmic complexity, providing diagnostic coverage across both the classical arithmetic and quantum correlational aspects of the computation.
\section{Related Work}
\label{sec:related_work}

Quantum debugging sits within a broader testing and verification ecosystem. Early studies highlighted challenges arising from probabilistic execution, limited observability, and hybrid quantum-classical workflows~\cite{miranskyy2021testing}, while more recent surveys highlight open problems in testing, assertions, and debugging methodologies~\cite{leite2025testing}. Empirical investigations show that developers still rely heavily on classical debugging practices and frequently encounter toolchain, transpilation, and SDK-related defects~\cite{zappin2025challenges,yousuf2025characterizing}. While not directly related to debugging, advances in verification paradigms, symbolic execution, and formal verification have enabled stronger program analysis, particularly for quantum error-correcting programs~\cite{choudhury2026verification,fang2024symbolic,huang2025efficient}. Together, these efforts highlight the need for debugging logic bugs in quantum computing.

 Statistical assertions first introduced hypothesis-testing-based checks for classical states, superposition, and entanglement in NISQ/logical quantum programs~\cite{huang2019statistical}. Subsequent work developed projection-based runtime assertions~\cite{li2020projection}, dynamic ancilla-assisted assertion circuits~\cite{liu2020quantum}, and systematic frameworks for precise and approximate state verification \cite{liu2021systematic}. Later studies connected assertions to error mitigation and expanded support to new environments and observables, including CUDA-Q and Bloch-vector assertions~\cite{li2022exploiting,li2025statistical,oldfield2025bloch}. Surveys identify assertion-based debugging as one of the most mature directions in quantum debugging, while noting ongoing challenges in scalability to FTQC platforms~\cite{kole2026assertion,rosa2026schrodinger}.

Beyond assertion implementations, recent work has also focused on integrating assertions into practical debugging workflows in current quantum settings. Debugging frameworks combine assertions with diagnostic procedures for fault localization~\cite{rovara2025framework}, while automatic assertion refinement improves debugging effectiveness by strategically inserting or relocating checks~\cite{rovara2025automatically}. Other work studies efficient assertion evaluation on quantum hardware under realistic noise and resource constraints~\cite{rovara2025framework2}. Automated fault localization, circuit analysis, property-based testing, and delta debugging have also enabled semi-automated diagnosis and explanation~\cite{maidl2025towards,pontolillo2026property}. Tool-oriented systems such as AssertsQ integrations further improve developer usability~\cite{veganzones2025assertsq}. 

Our work builds on the groundwork laid by these assertion-based debugging works, extending them to enable efficient implementation of FTQC systems.

\section{Conclusion}
\label{sec:conclusion}

In this work, we presented \sol{}, a systems-level debugging framework for FTQC applications that enables lightweight assertion-based debugging directly within encoded execution flows. \sol{} introduces practical debugging primitives that minimize spatial and temporal overhead while remaining compatible with evolving FTQC software and hardware ecosystems. Across multiple QEC code families, code distances, and encoded workloads, our evaluation demonstrates that practical FTQC debugging can be achieved with a low additional circuit-depth overhead (1.8\% on average). We believe this work represents an important first step toward practical software reliability infrastructures for large-scale fault-tolerant quantum computing systems.

\begin{acks}

We thank the anonymous ATC reviewers and the shepherd for their insightful feedback, which helped improve this work. This work was performed at Rice University and the Oak Ridge National Laboratory. It was supported by the U.S. Department of Energy, Office of Science, National Quantum Information Science Research Centers, Quantum Science Center. This manuscript has been authored in part by UT-Battelle, LLC under Contract No. DE-AC05-00OR22725 with the U.S. Department of Energy. The United States Government retains, and the publisher, by accepting the article for publication, acknowledges that the United States Government retains a non-exclusive, paid-up, irrevocable, worldwide license to publish or reproduce the published form of this manuscript, or allow others to do so, for United States Government purposes. The Department of Energy will provide public access to these results of federally sponsored research in accordance with the DOE Public Access Plan (http://energy.gov/downloads/doe-public-access-plan). We note the use of AI tools such as Grammarly for text editing and revision, and Claude Code (Opus 4.6) for programming assistance. The authors reviewed and verified all AI-generated content for correctness.

\end{acks}

\balance

\bibliographystyle{ACM-Reference-Format}
\bibliography{main}

\end{document}